# Acceleration AI Ethics,
# the Debate between Innovation and Safety,
# and Stability AI's Diffusion versus OpenAI's Dall-E

**James Brusseau**

Philosophy Department, Pace University, New York City

& Department of Information Engineering and Computer Science, University of Trento, Italy

## Abstract

One objection to conventional AI ethics is that it slows innovation. This presentation responds by reconfiguring ethics as an innovation accelerator. The critical elements develop from a contrast between Stability AI's Diffusion and OpenAI's Dall-E. By analyzing the divergent values underlying their opposed strategies for development and deployment, five conceptions are identified as common to acceleration ethics. Uncertainty is understood as positive and encouraging, rather than discouraging. Innovation is conceived as intrinsically valuable, instead of worthwhile only as mediated by social effects. AI problems are solved by more AI, not less. Permissions and restrictions governing AI emerge from a decentralized process, instead of a unified authority. The work of ethics is embedded in AI development and application, instead of functioning from outside. Together, these attitudes and practices remake ethics as provoking rather than restraining artificial intelligence.

## Introduction

The future of AI will not be decided by ethical conclusions, it will be decided by which ethics does the concluding. One proxy for deciding is Stability AI's Diffusion against OpenAI's Dall-E. While the two platforms share a talent for converting textual prompts into images, they divide along the edge of a threatening letter from a San Francisco congresswoman:

> Stability AI's Diffusion model is available for anyone to use without hard restrictions. This means Diffusion can create images that OpenAI's DALL-E currently blocks, including propaganda, violent imagery, pornography, copyright violations, and disinformation and misinformation…. l urge using established powers to control the release of unsafe Al models.

She wants Diffusion stopped.

The superficial explanation is the specific pictures, but diverse people will always chaotically disagree about which ones are harmful, under what circumstances, and why. The deeper reason for stopping Stability's Diffusion is the model's release without hard restrictions. This is an argument about ethical *attitudes*. It is about how ethics is configured to perceive AI from the moment it enters the world. Independent of any images coming later on, are restrictions inherently right for AI, or not?

Answering will force a confrontation between today's mainstream approach to AI ethics, and an alternative that has been circulating on the edges of philosophy, but that is only now crossing through data and algorithms.

## Conventional AI ethics and the acceleration alternative

Brussels has not taken the global lead in defining AI ethics, the principles and frameworks designed there nearly <u>*are* the definition</u>. The <u>General Data Protection Regulation</u> (GDPR) and subsequently the <u>Ethics Guidelines for Trustworthy AI</u> and then the <u>AI Act</u> are bedrock documents. California's much-discussed <u>Consumer Privacy Act</u> imitates the GDPR, and New York City's less-discussed <u>rules for AI hiring</u> mimic the European approach to technological trustworthiness. More generally, anywhere the term "trustworthy" appears in relation to AI, the implicit reference is the European standards which can be summarized, crudely but accurately, by a single sentence from one of the <u>authors,</u> "It is time to <u>make haste slowly</u> in the development of AI."

It may be that time. Certainly that is the expectation of OpenAI which has embraced the hesitations of today's established ethics to justify the incremental release Dall-E's increasingly powerful versions. It's impossible be certain, though, without considering the most penetrating criticism applied to the established standards: they are self-defeating. In the real world, making haste only slowly ends up postponing and <u>diminishing innovation</u>. More AI ethics, consequently, means less AI development until nothing remains to be hesitant about.

Reaching the terminal extreme seems unlikely, but the theoretical possibility opens the question about alternatives, and one – acceleration ethics for artificial intelligence – is represented by Stability's Diffusion. It is captured by these elements:

- Uncertainty is encouraging
- Innovation is intrinsically valuable
- The only way out is through: more, faster
- Decentralization
- Embedding

Each element responds to this question: *How can AI ethics be realigned to promote innovation?*

## Uncertainty is encouraging

An experiment can be done. The text of the letter lauding OpenAI and demanding restrictions on Stability can be copy/pasted into the respective text-to-image generation platforms. The results are telling. OpenAI returned a puppy and kitten picture, along with the message, "It looks like this request may not follow our content policy." Stability produced an echoing image that is curious, ghostly, and ultimately meaningless.

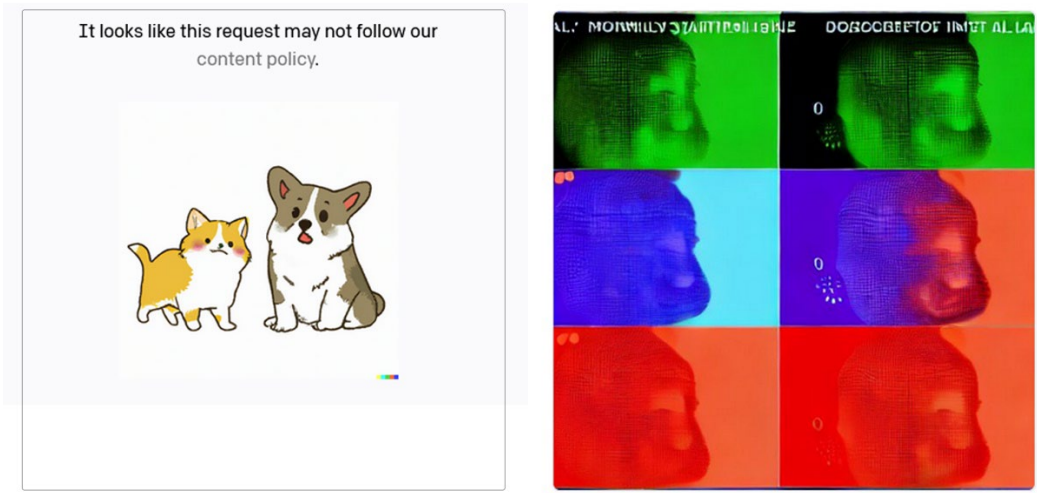

The absence of artistic depth does not obscure the implications of the distinct ways the two models responded to uncertainty, specifically, the uncertainty about what could happen if the image generators are turned loose in the world. For OpenAI – and conventional ethics – incomplete knowledge about AI's future effects is revealed as intrinsically negative. It is a reason to hesitate, a fear something might go wrong. That's why the inputted text was rejected, not because of a demonstrable harm that a produced image had actually done, instead because of what the graphic version might do, because the text-inspired image *may* violate policy.

This rationale of hesitation is explicit in a *New York Times* commentary on OpenAI:

> When I think about the potential for both good and harm that AI can do, the best approach is to parcel it out little by little. When I look at what OpenAI has done, that feels responsible to me.

No doubt it is responsible. It just needs to be underlined that what the term "responsible" means depends upon the ethics it serves, and this version of AI ethics postulates uncertainty as discouraging, as a good reason to limit the model's dissemination to "little by little."

Acceleration AI ethics conceives incomplete foreknowledge as positive and encouraging. It is unforeseeable possibilities and the incalculable potential of what is not yet determined. Instead of the unknown being daunting because there is not enough certainty, it is now provocative because there is so much uncertainty. The fact that we do not know what kinds of graphics will be produced, by whom, and for which purposes no longer requires limitations on generative AI, instead, it *is* the justification for allowing texts to bypass restrictions and convert into pixels.

On the ground of real users, the attitude toward uncertainty animated Stability to open its generative AI beyond selected engineers and pre-screened illustrators. Texts were accepted from nearly any project and interest: fashion design, architecture, civil engineering, chemistry, basement inventing. Already we see advanced image-generation being repurposed to aid robot navigation, to enable 3D construction planning, to facilitate the biochemical design of proteins. These outcomes are reassuring. But, they are no more than that. What is *emboldening* – and what spreads generative AI open for unforeseen use – is the allure of uncertainty. Acceleration ethics means forging ahead with AI generation *because* the future effects in the world are unknown, which is distinct from advancing despite not knowing.

Finally, in our own lives we have all experienced the attraction of pure potential. As one example, there is true digital nomadism, which is the travels of those who buy one-way airplane tickets to unfamiliar places because they want to change and be changed, without knowing how or why beforehand.

## Innovation is intrinsically valuable

The second difference between conventional and acceleration AI ethics is the value of innovation. Conventionally, any particular innovation is understood as possibly causing benefit *or* harm in society. The practical consequence is a familiar warning: just because we can build an AI tool does not mean we should. Instead, building should only come after projecting a sense of possible uses and abuses, benefits and harms, and then weighing to determine whether the innovation is worth having.

For acceleration AI ethics, innovation is *intrinsically* valuable: before considering any downstream effects in society, breakthroughs exist within their own sphere as innately worth doing and having. In this limited way, AI resembles art: both are creative and innovative and therefore worthwhile even before the finished work is publicly revealed. Consequently, the fact that an inventive generative model *can* be built converts into an ethical reason to do so.

One way to measure this split conception of innovation's value is through the burden of restriction. Conventionally, AI designers have had the burden of showing why their projects should not be restricted. Distinctly and for acceleration, the default setting is develop and deploy, and the burden shifts to those who want to stop it. OpenAI and Stability incarnate this difference almost perfectly. To access its model, OpenAI requires an extensive sign-in process, personal information confirmation, and also passing a miscreant filter. The developers allow their machine out into the world only as far as *they* can justify because the burden is on them to show they are being responsible. Contrastingly, Stability allows anyone to dive straight in and generate images. No name required, no verification of an email address or confirmed approval. In ethical terms, the principle is to develop models and facilitate broad use, and that should go on until persuasive reasons are elaborated for why it should not. Because innovation is intrinsically valuable, the burden is on users and others to restrain the innovators.

For conventional AI ethics, we make haste slowly by confirming an innovation's social benefits before advancing. For acceleration, innovation's value is intrinsic, which propels advances before subsequent effects can be measured. This does not make AI unstoppable, it only requires ethical reasons for halting to overcome the creative reason for doing AI in the first place.

## The only way out is through: more, faster

The third difference between standard and acceleration AI ethics is the structure of response to problems rising from innovation. When generative image models produce deepfake pornography, or when language models spew crude racism, the conventional reaction is to slow the AI. For its image generator, OpenAI applied the brakes by initially limiting users to 400. That allowed the company to humanly review the pictures being composed and study potential misuse. While that reviewing and studying was happening, research was pausing. Which is an iterative approach, according to OpenAI, meaning engineers can take the next step forward only after the last one has been confirmed solid and safe.

At the extreme, hesitating progress grinds to a halt. When Microsoft's generative AI experiment in language got caught spouting offensive rejoinders, it was pulled offline and shelved forever. No doubt parts of the project are being resuscitated, but that doesn't change the underlying reality:

conventional ethics responds to problematic innovation with diminished compute and restricted accessibility.

For acceleration ethics, the response is more, faster: AI straight ahead until coming through on the other side. Stability exemplifies, initially because its Diffusion manifestly qualifies as problematic innovation. The model produces captivating graphics reflecting the minds and words of even the most perverted and rabid. There is no denying the potentially noxious uses. The reaction, however, is not to slow technological development and limit participants, instead, Stability accelerates innovation and broadens the community of engineers and users. One method: a $200,000 prize offered to anyone for the best open-source deepfake detector. That allowed Diffusion to run unabated while also stimulating a parallel field of AI research and innovation to contain Diffusion toxicity. Technological abuses, in other words, are met with amplified technology. The same mentality held at MIT's Madry Lab where a team of machine learning students jumped into the problem and produced a data shield for images. It essentially short-circuits generative algorithms with clever pixel effects, and the result is a mathematical coating against deepfakery. Of course, there is a distance between academic experiments and worldly effects, and it is also true that prizes may not guarantee anti-abuse technology, but these are empirical questions, not ethical ones.

The applied ethics – the *kind* of ethics that will be applied – starts from the recognition that generative AI will create benefits and harms graphically. From there, the dispute opens about the direction and speed of the response: Back and slower, or ahead and faster? Is the answer a limited group of 400 users minting images while engineers cautiously refine their algorithms with Trust and Safety officers peering over their shoulders. Or, is it wildcat coders chasing $200,000 prizes, and motley graduate students whose minds are unpolluted by accepted wisdom? Like all true dilemmas, the answer will only come when it is too late.

For now, the conventional response to innovation-caused problems is to make haste slowly, one firmly grounded step before another. The acceleration response resembles big wave surfing in this sense: maybe it would have been better to not catch the wave in the first place, but once you are on the board and the water is curling above, there is only one way to deliverance.

## Decentralization

The *Ethics Guidelines for Trustworthy AI* was produced by a set of authors literally titled the "High-Level Expert Group," which could be a universal motto for centralization. OpenAI's internal ethics team gathers under the humbler designation of Trust and Safety, but regardless of the words, the idea is to issue guidelines directing subsequent development and applications.

There are two vectors, consequently, to AI ethics centralization. Guidelines come before AI applications get publicly released and, distinctly, a narrow group develops the guidelines for the wider community. Decentralized ethics disrupts both. Instead of guidelines coming before and conditioning AI use, the AI use subsequently produces the guidelines. And, instead of the few deciding for the many, the many decide for each other.

In an interview with the *New York Times*, Stability's founder provided a curious example:

> If you're a centralized thing what they do is they randomly allocate gender if it's a non-gendered word. So you have female Sumo wrestlers, if you type in Sumo wrestler.

The confusion starts with gender uncertainty in nouns. Image generators process the text prompt "waitress" by directly assigning a feminine shape, but if the request is "wrestler," there are

5CEPE 2023: International Conference on Computer Ethics: Philosophical Enquiry 2023
Illinois Institute of Technology, Chicago, IL, United States.

questions. The visible one asks about male or female, while underneath there is: *How will the decision be made?*

For centralized ethics, the decision logic is well established. First there is the designation of the regulatory group, then the promulgation of permissions and prohibitions, then developers build their model within the established boundaries and, finally, the model is released to users who begin generating images. In the wrestling case, the critical moment came when centralized regulators prohibited imbalanced representations of gender-neutral nouns (representational fairness). Developers responded efficiently by injecting gender randomness into the model *before* release to the broader community of human users, and the result was visions of female Sumo wrestlers.

Decentralization reverses the sequence of guidelines and deployment, while also shifting guideline decisions toward the broad community of users. Consequently, answers to the question about gendering images – or to questions about permissions and prohibitions generally – lead to the images that users have *already* generated. These answers are recursive in the digital sense that they have no localizable beginning: it is impossible to say how the word and image associations got started. Once they are going, however, users' prompts, and then their honing modifications begin to reshape the model's functioning.

There is an associated technical jargon – reinforcement learning with human feedback – but on the conceptual level the ethical operation is straightforward. Initial textual prompts typed into the generative AI produce satisfying or unsatisfying images. When unsatisfying, the text inputs are modified: "Sumo" is re-submitted as "male Sumo" or "female Sumo," depending on the user. As users multiply, and their new image generations proliferate, the gender correspondences recycle into the data pool used to create and shape the next round of graphics. While that happens, any visible gendering originally dictated by centralized guidelines or algorithmic inclinations begins giving way to the broad community of users and their projects. And that *process* creates effective ethical guidelines for gender in subsequent Sumo wrestler image productions. So, instead of guidelines and then produced graphics, the produced graphics *are* the guidelines for future production.

Ideally, as the reinforcement learning grinds forward, the gender predictions will begin adhering to specific creators' personal intentions even without refining and detailing their prompts: users will type "Sumo" and get what they individually intended without further specification. When that happens, gender ethics in generative models has decentralized.

Decentralization does not equal accuracy. Image generators governed exclusively by past text-image regularities have responded to the word "swan" with a black version, presumably because the internet actually includes more pictured references to black swans than to the common variety.

Regardless, centralized and decentralized approaches are both vulnerable to their own patterns of failure, but the specific breakdowns – Sumos are female, swans are black – are transitory. What endures is the underlying ethics of permissions and restrictions governing the connections between words and graphic things. Either it comes before and from the few, or after and from many.

## Embedded

Centralized ethicists – members of high-level expert groups – issue restrictions from above and outside the technology. Sometimes these are called guidelines, other times principles, or rules, or regulations. In any case, the role is more akin to policing than to participation. At the extreme, constraining answers are provided even before any questions can be asked.

Contrastingly, when ethics emerges from the model's development and application, ethicists are naturally embedded in the process. Their role is to collaborate with engineers and users to spin the language and concepts of human values out of information and algorithms. The skill is closer to translation than to restriction, closer to exploration than control. For those working inside the machine, formulating AI dilemmas is more imperative than resolving them, and provocative questions will be asked even when no answers exist.

## Conclusion

If the future of AI will be decided not by ethical decisions but by which ethics does the deciding, then where we are headed is already indicated by the divergence between OpenAI and Stability. On the conventional side there is the ideal of making haste slowly. For acceleration, distinguishing elements include:

- Uncertainty is encouraging. The unknown displays potential more than risk, and therefore AI models are developed for the same reason that digital-nomads travel: *because* we do not know how we will change and be changed.
- Innovation is intrinsically valuable. Because it resembles artistic creativity in being worthwhile even before considering social implications, an ethical burden tips: engineers no longer need to justify starting their models, instead, others need to demonstrate reasons for halting.
- The only way out is through: more, faster. When ethical problems arise, the response is not to slow AI and veer away, it is straight ahead until emerging on the other side. Accelerating AI solves AI problems.
- Decentralization. The overarching ethics of permissions and restrictions governing AI derive from the broad community of users and their uses, instead of representing a select group's preemptory judgments.
- Embedding. Ethics works inside AI and with engineers to pose questions about human values, instead of remaining outside and emitting restrictions. Formulating AI dilemmas is more imperative than resolving them.

## Declaration

Author declares there is no conflict of interest or external funding.

## References embedded in text